\documentclass[twocolumn,prd,nofootinbib]{revtex4-1}

\usepackage{graphicx}
\usepackage{epstopdf}
\usepackage{dcolumn}
\usepackage{bm}
\usepackage{hyperref}
\usepackage{color}
\usepackage{mathrsfs}
\usepackage{fancyhdr}
\usepackage{amsmath,amsfonts,amssymb}
\usepackage[normalem]{ulem}	
\usepackage{capt-of}

\newcommand\be{\begin{equation}}
\newcommand\ee{\end{equation}}
\newcommand\ba{\begin{eqnarray}}
\newcommand\ea{\end{eqnarray}}

\newcommand\doth{\dot{h}}
\newcommand\sQ{{\sigma_{Q}^2}}
\newcommand\nn{\nonumber}
\newcommand\lb{\label}

\newcommand\ham{{\cal H}}
\newcommand\mub{\bar{\mu}}

\newcommand{\eq}[1]{Eq. (\ref{#1})}
\newcommand{\eqs}[2]{Eqs. (\ref{#1}) - (\ref{#2})}

\definecolor{gre}{rgb}{0,0.6,0}
\definecolor{red}{rgb}{1,0,0}
\definecolor{blu}{rgb}{0,0,1}

\begin{document}

\title{Exhaustive investigation of the duration of inflation in effective anisotropic loop quantum cosmology}

\author{Linda Linsefors}%
 \email{linsefors@lpsc.in2p3.fr}
\affiliation{%
LPSC, Université Grenoble-Alpes, CNRS/IN2P3\\
53, avenue des Martyrs, 38026 Grenoble cedex, France
}%

\author{Aurelien Barrau}%
 \email{Aurelien.Barrau@cern.ch}
\affiliation{%
LPSC, Université Grenoble-Alpes, CNRS/IN2P3\\
53,avenue des Martyrs, 38026 Grenoble cedex, France
}

\date{\today}

\begin{abstract}
This article addresses the issue of estimating the duration in inflation in bouncing cosmology when anisotropies, inevitably playing an important role, are taken into account. It is shown that in Bianchi-I loop quantum cosmology, the higher the shear, the shorter the period of inflation. In a range of parameters, the probability distribution function of the duration of inflation is however peaked at values compatible with data, but not much higher. This makes the whole bounce/inflationary scenario consistent and phenomenologically appealing as all the information from the bounce might then not have been fully washed out.
\end{abstract}


\maketitle

\section{Introduction}

Loop quantum gravity (LQG) is an attempt of deriving a background-independent  and non-perturbative quantization
of general relativity. In the canonical formalism, it relies on Ashtekar variables, namely SU(2) valued connections and  conjugate densitized triads. The quantization is basically obtained through holonomies of the connections and fluxes of the densitized triads. Quite a lot of impressive results were recently obtained including an extensive covariant reformulation (see, {\it \emph{e.g.}}, \cite{rovelli1} for introductions and reviews). 

Loop quantum cosmology (LQC) tries to capture the main features of LQG
in the highly symmetric situations relevant for cosmology. A generic and important result of LQC is that the big bang is replaced by a big bounce due to strong repulsive quantum geometrical effects (see, {\it \emph{e.g.}}, \cite{lqc_review} for reviews and \cite{diener} for a recent numerical test of the bounce robustness).\\

In all bouncing cosmologies (except ekpyrotic models build to avoid this problem), either from the loop approach or any other, the issue of anisotropies is crucial for a very clear reason: the shear term basically varies as $1/a^6$ where $a$ is the ``mean'' scale factor of the Universe. When the Universe is contracting, the shear term becomes more and more important and might drive the dynamics. The very reason why the shear can be safely neglected in standard inflationary cosmology is precisely the reason why it becomes essential in bouncing models. When one assumes spatial homogeneity, anisotropic hypersurfaces admit transitive groups of motion that must be three- or four-parameters isometry groups. The four-parameters groups having no simply transitive subgroups, they will not be considered here. There are nine algebraically nonequivalent three-parameters simply transitive Lie groups: Bianchi I through IX, all with well known structure constants. The flat, closed and open generalizations of the FLRW model are respectively called Bianchi-I, Bianchi-IX and Bianchi-V. In the following, we focus on the Bianchi-I model to study the dynamics around the bounce. This is meaningful as the Universe is nearly flat today and as the relative weight of the curvature term in the Friedmann equation is decreasing with increasing values of the density.\\

Different studies have already been devoted to Bianchi-I LQC \cite{mubar,bianchiI_lqc}. In particular, it was shown that the bounce prediction resists the introduction of anisotropies. 
As the main features of isotropic LQC are well captured by semi-classical effective equations \cite{diener}, we will assume that this remains true when anisotropies are included. 

In a previous work, we have systematically explored the full solution space in a way that has not been considered before and have derived the LQC-modified generalized Friedmann equation that was still missing \cite{Linsefors:2013bua}. In another work, we established that, if initial conditions are set in the contracting phase, the duration of inflation is not anymore a free parameter but becomes a highly peaked distribution in isotropic effective LQC \cite{Linsefors:2013cd}. In this article, we address both questions at the same time: how is the duration of inflation affected by anisotropies that {\it must} be taken into account in any consistent bouncing approach?

The solutions of effective equations of Bianchi-I LQC, and the effect of the shear on slow-roll inflation, were previously studied in \cite{Gupt:2013swa}. The main difference between this article and \cite{Gupt:2013swa} is that we set the initial conditions in the far past, well before the bounce, while \cite{Gupt:2013swa} set their initial conditions at the bounce. This difference is further discussed in Section \ref{dis}.

\section{Gravitational sector}
Except  the definition given in \eq{s}, this section is mostly a summary of the first half of \cite{Linsefors:2013bua}.\\

The metric for a Bianchi-I spacetime is given by
\be\lb{ds}
ds^2 = -N^2 d\tau^2 + a_1^2dx^2 + a_2^2dy^2 + a_3^2dz^2,
\ee
where the $a_i$ denote the directional scale factors. The classical gravitational hamiltonian is
\be
\ham_G = \frac{N}{\kappa\gamma^2}\left(\sqrt{\frac{p_1 p_2}{p_3}}c_1c_2+\sqrt{\frac{p_2 p_3}{p_1}}c_2c_2+\sqrt{\frac{p_3 p_1}{p_2}}c_2c_3\right),
\label{classhamG}
\ee
with Poisson brackets
\be
\{c_i,p_j\}=\kappa\gamma\delta_{ij},
\ee
where $\gamma$ is the Barbero-Immirzi parameter.

The classical directional scale factors can be calculated from $p_i$,
\be
a_1 = \sqrt{\frac{p_2 p_3}{p_1}}
\qquad \text{and cyclic expressions.}
\ee
The generalized Friedmann equation is
\be
H^2=\sigma^2+\frac{\kappa}{3}\rho,
\label{clfried}
\ee
where
\be
H:=\frac{\dot{a}}{a}=\frac{1}{3}(H_1+H_2+H_3)~,~a:=(a_1a_2a_3)^{1/3},
\ee
\be\lb{clShear}
\sigma^2=\frac{1}{18}\Big( (H_1-H_2)^2+(H_2-H_3)^2+(H_3-H_1)^2 \Big),
\ee
\be
H_1:=\frac{\dot{a}_1}{a_1}=
-\frac{\dot{p}_1}{2p_1}+\frac{\dot{p}_3}{2p_3}+\frac{\dot{p}_3}{2p_3}
\qquad \text{and cyclic expressions.}
\ee

To account for LQC effects, the holonomy correction has to be implemented. It is rooted in the fact that the Ashtekar connection cannot be promoted to be an operator whereas its holonomy can. It is assumed to capture relevant quantum effects at the semi-classical level. Following the usual prescription, one substitutes
\be
c_i\rightarrow \frac{\sin(\mub_i c_i)}{\mub_i}
\ee
in the classical Hamiltonian. The $\mub_i$ are given by
\be
\mub_1 = \lambda\sqrt{\frac{p_1}{p_2p_3}}
\qquad \text{and cyclic expressions,}
\ee
where $\lambda$ is the square root of the minimum area eigenvalue of the LQG area operator ($\lambda=\sqrt{\Delta}$) \cite{mubar}. The quantum corrected gravitational Hamiltonian is:
\begin{widetext}
\be\lb{ham}
\ham_G=-\frac{N\sqrt{p_1p_2p_3}}{\kappa\ \gamma^2\lambda^2}\Big[\sin(\mub_1c_1)\sin(\mub_2c_2)+\sin(\mub_2c_2)\sin(\mub_3c_3)+\sin(\mub_3c_3)\sin(\mub_1c_1)\Big].
\ee

It can be shown that all gravitational observables are fully described by the $h_i$: 
\be\lb{hi}
h_1:=\mub_1c_1=\lambda\sqrt{\frac{p_1}{p_2p_3}}c_1
\qquad \text{and cyclic expressions.}
\ee
As it could be expected, the initial six degrees of freedom $(c_i,p_i)$ reduce to only three physical degrees of freedom $h_i$. This is because only the $H_i$ and not the $a_i$ are observables. The $a_i$ can indeed be changed by a rescaling of the corresponding coordinate, while $H_i$ remains invariant under any coordinate transformation that preserves the form of the metric given by \eq{ds}.

The equations of motion for the $h_i$ are derived from Eqs. (\ref{ham}) and (\ref{hi}):
\begin{multline}
\doth_1=\frac{1}{2\gamma\lambda}\Big[
(h_2-h_1)(\sin h_1+\sin h_3)\cos h_2 +
(h_3-h_1)(\sin h_1+\sin h_2)\cos h_3\Big] -
\frac{\kappa\gamma\lambda}{2}(\rho+P)
\\ \text{and cyclic expressions,}
\label{doth}
\end{multline}
where $\rho$ is the total matter energy density and $P$ is the total pressure.

The average and directional Hubble parameters can now be re-expressed in terms of $h_i$:
\be
H_1:=\frac{\dot{a}_1}{a_1}=\frac{1}{2\gamma\lambda}\Big[\sin(h_1-h_2)+\sin(h_1-h_3)+\sin(h_2+h_3)\Big]
\qquad \text{and cyclic expressions,}
\label{h}
\ee
and
\be
H:=\frac{\dot{a}}{a}=\frac{1}{3}(H_1+H_2+H_3)=\frac{1}{6\gamma\lambda}\Big[\sin(h_1+h_2)+\sin(h_2+h_3)+\sin(h_3+h_1)\Big].
\label{H}
\ee
The LQC-modified generalized Friedmann equation can now be written: 
\be\lb{fried}
H^2=\sQ+\frac{\kappa}{3}\rho-\lambda^2\gamma^2\left(\frac{3}{2}\sQ+\frac{\kappa}{3}\rho\right)^2,
\ee
where we define the quantum shear $\sQ$ as
\be\lb{sQ}
\sQ:=\frac{1}{3\gamma^2\lambda^2}\left(1-\frac{1}{3}\Big[\cos(h_1-h_2)+\cos(h_2-h_3)+\cos(h_3-h_1) \Big]\right).
\ee
\end{widetext}
The LQC-modified generalized Friedman equation given by \eq{fried} puts upper limits on both $\sQ$ and $\rho$.
\be
\sQ\leq\sQ_c := \frac{3}{2\lambda^2\gamma^2}
\quad,\quad
\rho\leq\rho_c := \frac{\kappa}{3\lambda^2\gamma^2}.
\ee
It should be noticed that $\rho_c$ is the same as in the isotropic case.

In the limit $h_i\rightarrow n\pi$ with the same integer $n$ for all $i$, then $\sQ\approx\sigma^2$. A special case of this is of course $\lambda\rightarrow 0 \Rightarrow h_i\rightarrow 0$.

From Eqs. (\ref{doth}) and (\ref{H}), one finds that 
\be
(\doth_i-\doth_j)=-3H(h_i-h_j),
\ee
and from this, it follows that
\be\lb{prop}
(h_i-h_j)\propto a^{-3}.
\ee

Without any loss of generality, one can choose the labeling of the spatial dimensions such that initially 
\be\lb{leq}
h_1\leq h_2\leq h_3.
\ee
Because of Eq. (\ref{prop}), we know that this inequality will hold at any time.\\

We now define a the symmetry variable for the anisotropy.
\be\lb{s}
S:=\frac{(h_2-h_1)-(h_3-h_2)}{(h_3-h_1)}.
\ee
It follows form Eq. (\ref{prop}), that $S$ is a constant of motion, and it follows form its definition and Eq. (\ref{leq}) that $-1\leq S\leq1$.

The classical shear $\sigma^2$ can in principle be calculated from $H$, $\sQ$ and $S$, since they contain all the physical information about the gravitational sector. However the exact relation is rather complicated and outside the scope of this article.

\section{Matter}\lb{M}
A matter field has to be introduced. It will of course play a crucial role in the inflationary phase. We will choose a simple massive scalar field with potential
\be
V(\phi)=\frac{m^2\phi^2}{2}.
\ee
This is the simplest potential that will generate slow roll inflation, and therefore a good generic toy model. 

Given this potential, the equation of motion for the scalar field is:
\be\lb{eomMater}
\ddot{\phi}=-3H\dot{\phi}-m^2\phi.
\ee
The above equation, together with Eqs. (\ref{doth}), is the full set of equations of motion describing the system under study in this work.

\section{Early evolution}
As shown in \cite{Linsefors:2013bua}, there are many solutions that never reach a classical limit, neither in the past nor in the future. However, in this study, we restrict ourselves to solutions that behave like a contracting classical universe in the remote past. As it was also found in \cite{Linsefors:2013bua}, all these solutions, and only these solutions, will approach the behavior of a classical expanding universe in the future. 

For all the solutions of interest, there will therefore be a period in the far past where the solution behaves just like a classical contracting universe. This gives  
\be\lb{sQprop}
\sQ\propto a^{-6}.
\ee
The above relation can also be verified explicitly from \eq{sQ} and \eq{prop} in the limit $h_i\rightarrow n\pi$. For the solutions of interest, \eq{sQprop} holds when $\sQ\ll1$ in Planck units.

The matter content is slightly more complicated due to the oscillatory behavior of \eq{eomMater}. 
Under the conditions
\be\lb{cond}
\rho\ll\rho_c \ ,\ \ 
H<0 \ ,\ \
H^2 \ll m^2 \ ,\ \
\sQ\ll\frac{\kappa}{3}\rho\ ,
\ee
the solutions to \eq{eomMater} are well approximated by \cite{Linsefors:2013cd}
\be\lb{rhot}
\rho(t)=\rho_0\left(1-\frac{1}{2}\sqrt{3\kappa\rho_0}\left(t+\frac{1}{2m}\sin(2mt+2\delta)\right)\right)^{-2},
\ee
\be
m\phi(t)=\sqrt{2\rho(t)}\sin(mt+\delta),
\ee
and
\be\lb{phit}
\dot\phi(t)=\sqrt{2\rho(t)}\cos(mt+\delta),
\ee
for some parameters $\rho_0$ and $\delta$. Note that $\rho(0)$ is approximately, but not exactly, equal to $\rho_0$.

For the solutions of interest, the conditions given by \eq{cond} will always be met at some early enough time. \eq{cond} implies $\sQ\ll1$ so that \eq{sQprop} is also true if the conditions of \eq{cond} are fulfilled.\\

If we average out the matter oscillations we find that
\be\lb{rhoProp}
\left<\rho\right> \propto a^{-3},
\ee
where $\left<\cdot\right>$ is the time average over one oscillation, or equivalently the average over all angles $\delta$. At $t=0$ we have $\left<\rho(0)\right>=\rho_0$.  

Combining the above with \eq{sQprop} gives
\be\lb{prop2}
\sQ \propto \left<\rho\right>^2.
\ee

\section{Simulations}

We have performed exhaustive numerical simulations (with control of the numerical errors) to investigate the duration of inflation as a function of the different variables entering the dynamics. 
All the simulations carried out is this work are started in the contracting phase, with a small energy density and shear, so that everything is well described by unmodified classical equations. 

All simulations  start from initial conditions fulfilling \eq{cond}. As initial conditions for the matter sector, we use \eqs{rhot}{phit} with $t=0$:
\be\lb{rho0}
\rho(0)=\rho_0\left(1-\frac{1}{2}\sqrt{3\kappa\rho_0}\frac{1}{2m}\sin(2\delta)\right)^{-2},
\ee
\be
m\phi(0)=\sqrt{2\rho(0)}\sin(\delta),
\ee
and
\be\lb{phi0}
\dot\phi(0)=\sqrt{2\rho(0)}\cos(\delta).
\ee
As demonstrated in \cite{Linsefors:2013cd}, these initial conditions, together with a flat distribution of $\delta$, give a natural distribution over solutions with different ratios of kinetic to potential energy. This is an important point for this study.

Since everything is initially very small in Planck units, we can approximate:
\be
H(0)\approx\frac{1}{3\gamma\lambda}\Big[h_1(0)+h_2(0)+h_3(0)\Big],
\ee
\ba
\sQ(0)\approx\frac{1}{18\gamma^2\lambda^2}\Big[&\big(h_1(0)-h_2(0)\big)^2+\big(h_2(0)-h_3(0)\big)^2& \nn\\
&+\big(h_3(0)-h_1(0)\big)^2\Big].&
\ea
Solving the above equations together with Eq. (\ref{s}) gives for $h_1(0)$, $h_2(0)$ and $h_3(0)$:
\ba
h_1(0)&\approx&\gamma\lambda H(0)+\gamma\lambda\frac{-3-S}{\sqrt{3+S^2}}\sqrt{\sQ(0)}\lb{h10},\\
h_2(0)&\approx&\gamma\lambda H(0)+\gamma\lambda\frac{2S}{\sqrt{3+S^2}}\sqrt{\sQ(0)},\\
h_3(0)&\approx&\gamma\lambda H(0)+\gamma\lambda\frac{3-S}{\sqrt{3+S^2}}\sqrt{\sQ(0)}.
\ea
Finally, the initial conditions have to fulfill Eq. (\ref{fried}), and since we are in the contracting branch we get the negative solution for $H(0)$,
\be\lb{h0}
H(0)=-\sqrt{\sQ(0)+\frac{\kappa}{3}\rho(0)-\lambda^2\gamma^2\left(\frac{3}{2}\sQ(0)+\frac{\kappa}{3}\rho(0)\right)^2}.
\ee\\

The initial conditions of the simulations are now fully specified by Eqs. (\ref{rho0})-(\ref{phi0}) and (\ref{h10})-(\ref{h0}), and the parameters $\rho_0$, $\delta$, $\sQ(0)$ and $S$. 

To make sure that the approximations used for the initial conditions hold, the parameters must fulfill:
\be\lb{ini_cond}
\sQ(0)\ll\frac{\kappa}{3}\rho_0\ll m^2
\ee
in Planck units, together with
\be
-1\leq S\leq1 \quad \text{and}\quad 0\leq\delta<2\pi,
\ee 
which follows from the definitions.

We have chosen to keep $\rho_0$ fixed for all simulations in this article. Different proportions between matter and shear are achieved by varying $\sQ(0)$. Fixing $\rho_0$ have the disadvantage that we are limited in how high $\sQ$ can be chosen for that specific energy density, since we also have to respect \eq{ini_cond}. However, since we must anyway restrict ourselves to a finite parameter range, we just have to choose $\rho_0$ small enough to cover the solution space of interest.
In all simulations $\frac{\kappa}{3}\rho_0=10^{-3}m^2$, and $\sQ(0)\in[0,10^{-5}m^2]$.\\

For all the simulations in this work, we have used $m=1.21\times 10^{-6}$, as favored by observation \cite{linde}. In addition, we have used $\gamma=0.2375$ and $\lambda=\sqrt{4\sqrt{3}\pi\gamma}$ \cite{rovelli1} which gives $\rho_c=0.41$ and $\sQ_c=1.52$.

\begin{figure}
	\begin{center}
		\includegraphics[width=\columnwidth]{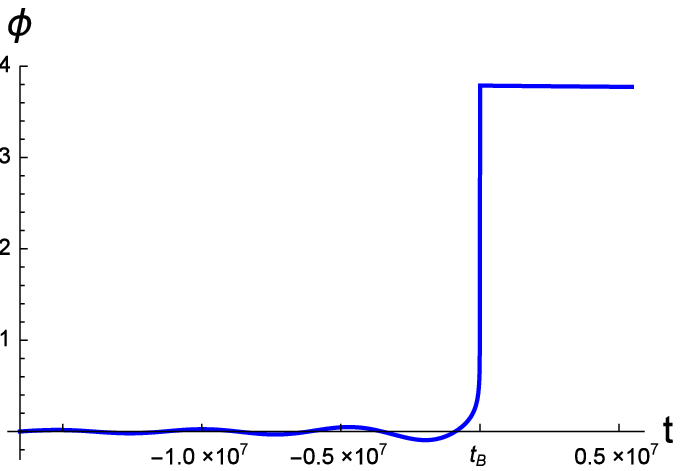}\\
		\includegraphics[width=\columnwidth]{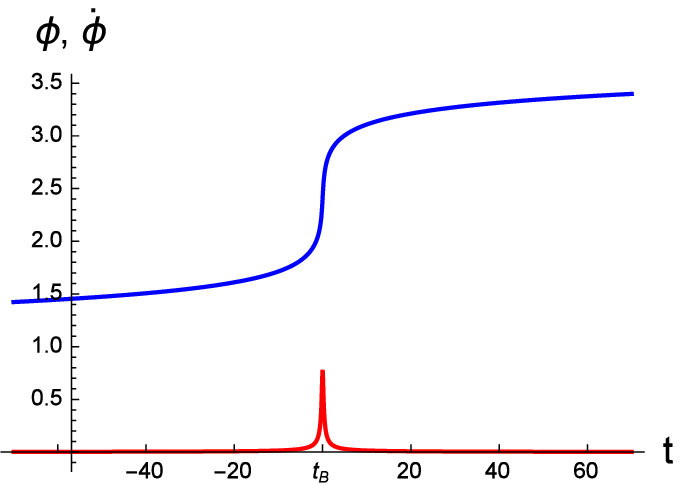}\\
		\includegraphics[width=\columnwidth]{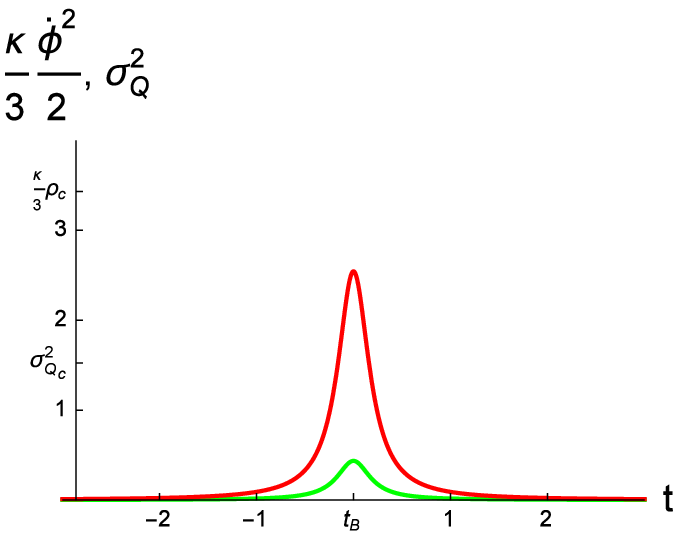}
		\caption{Typical evolution of $\phi$, $\dot{\phi}$, and $\sQ$ as a function of time. $t_B$ is the time of the bounce, when $H=0$. The blue line is $\phi$, the red is $\dot{\phi}$ or $\frac{\kappa}{3}\,\frac{\dot{\phi}^2}{2}$, and the green is $\sQ$. All the above plots are from the same simulation.
		Note that the kinetic energy is always completely dominating  over the potential energy at the bounce $\frac{\dot{\phi}_B^2}{2}\gg\frac{m^2\phi_B^2}{2}$. The fraction of shear and potential energy at the bounce varies, depending on the initial conditions. The initial conditions are $\sQ(0)=10^{-3}\frac{\kappa}{3}\rho_0$, $S=0$ and $\delta=0$.}
		\label{typ}
	\end{center}
\end{figure}

\begin{figure*}
\begin{minipage}{\columnwidth}	
		\includegraphics[width=0.48\columnwidth]{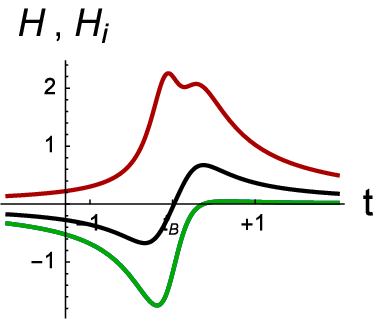}\hfill
		\includegraphics[width=0.48\columnwidth]{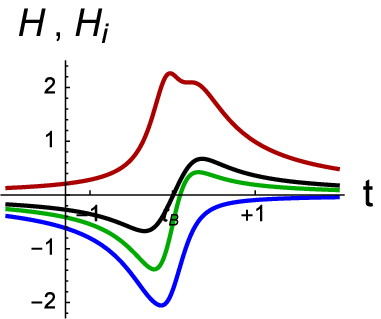}\\
		\includegraphics[width=0.48\columnwidth]{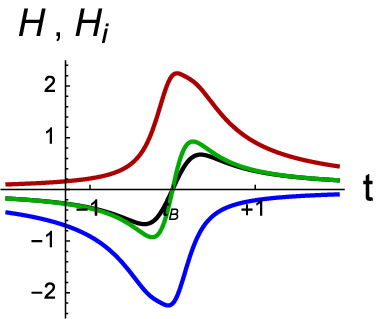}\\
		\includegraphics[width=0.48\columnwidth]{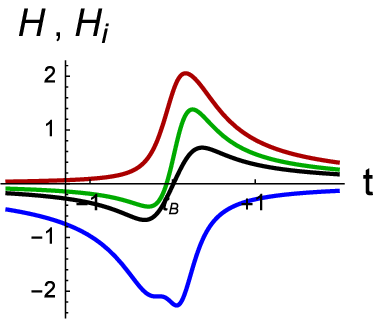}\hfill
		\includegraphics[width=0.48\columnwidth]{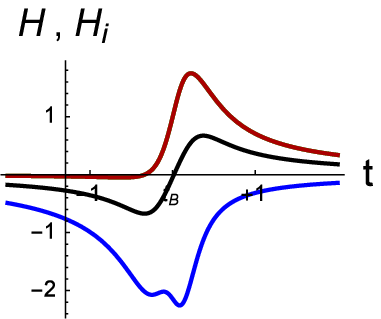}
		
		\captionof{figure}{
		Details of the bounce showed for different values of the symmetry factor $S$. Upper left: $S=-1$. Upper right: $S=-0.5$. Center: $S=0$. Lower left: $S=0.5$. Lower right: $S=1$. Black is $H$, red is $H_3$, green is $H_2$ and blue is $H_1$. $\sQ(0)=10^{-2}\frac{\kappa}{3}\rho_0$ and $\delta=0$ for all the above plots. In the upper left plot, the blue line is hidden by the green one. In the lower right plot, the green line is hidden by the red one.
		}
		\label{S}
\end{minipage}
\hfill
\begin{minipage}{\columnwidth}
		\includegraphics[width=\columnwidth]{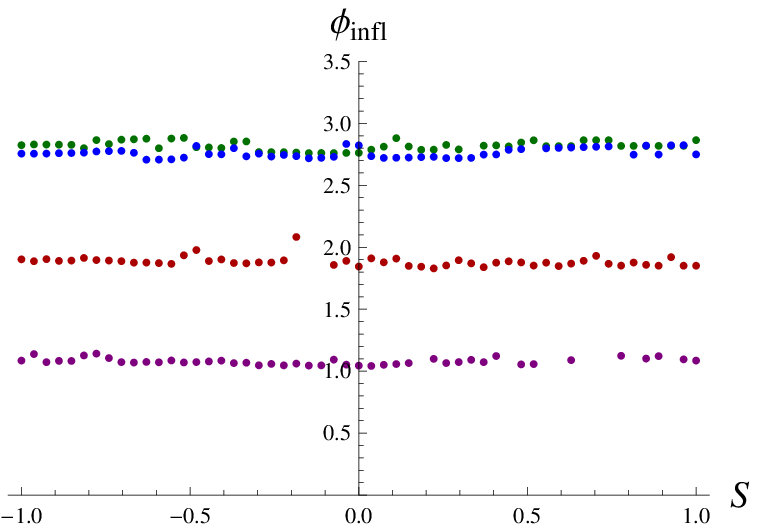}\\
		\includegraphics[width=\columnwidth]{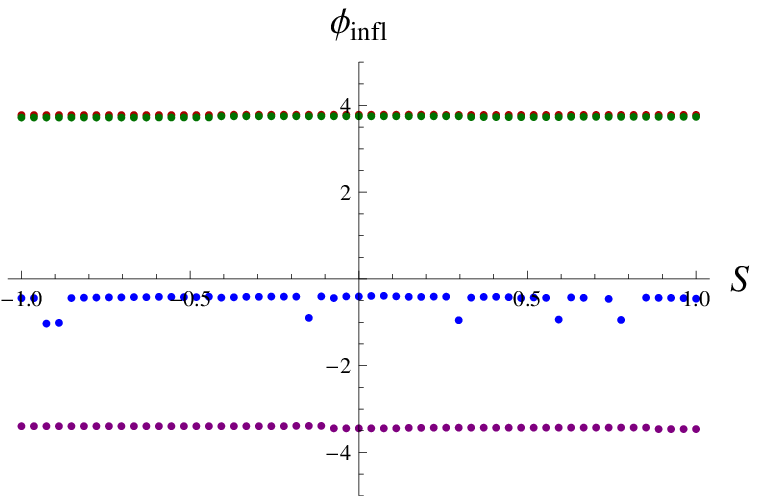} 
		\captionof{figure}{
		Scalar field at the beginning of inflation as a function of $S$. Upper: $\sQ(0)=10^{-2}\frac{\kappa}{3}\rho_0$. Lower: $\sQ(0)=10^{-3}\frac{\kappa}{3}\rho_0$. Red is $\delta=0$, green is $\delta=\frac{\pi}{4}$, blue is $\delta=\frac{\pi}{2}$ and purple is $\delta=\frac{3\pi}{4}$. Note that in the lower plot, the green points are right on top of the red ones and are therefore hard to be seen.
		}
		\label{SS}
\end{minipage}
\end{figure*}

%
		%


\begin{figure*}
	\begin{center}
		\includegraphics{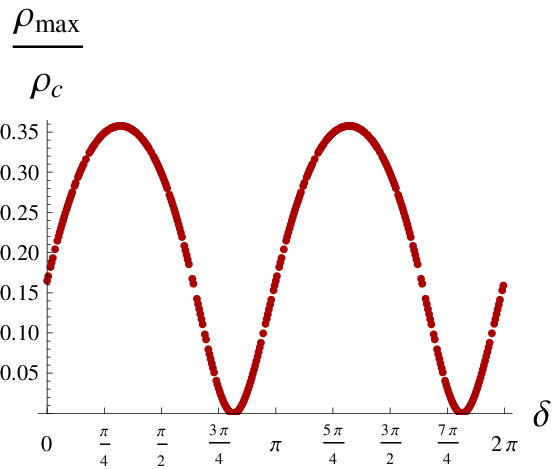}\hfill 
		\includegraphics{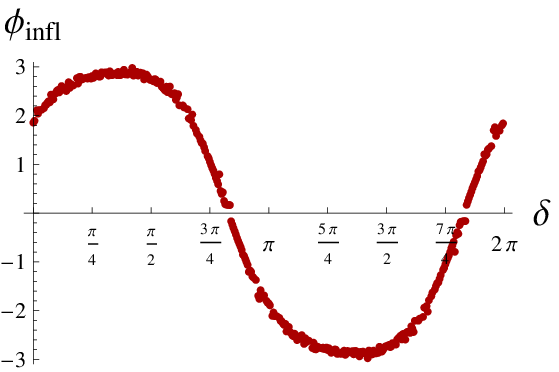}\hfill 
		\includegraphics{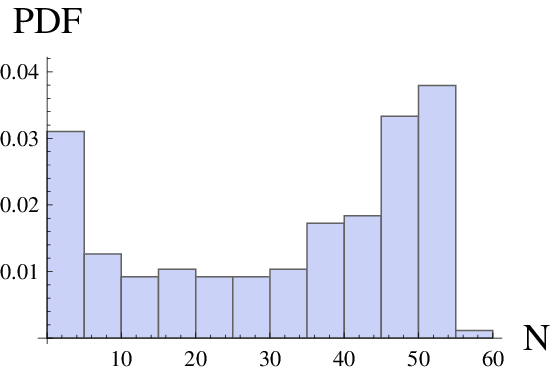}\\
		\includegraphics{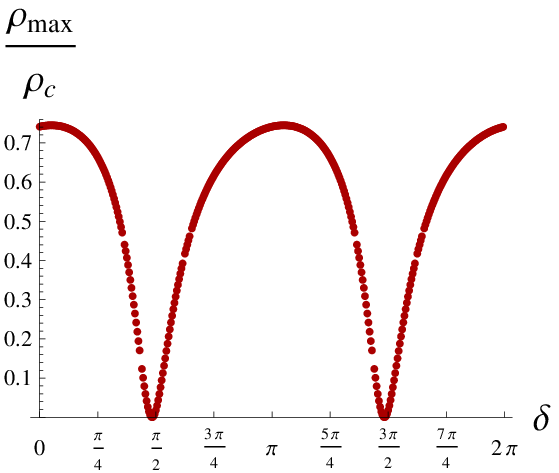}\hfill 
		\includegraphics{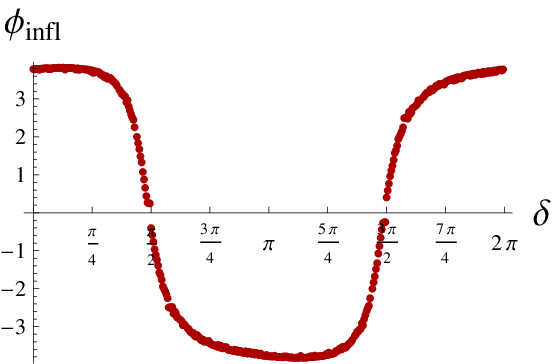}\hfill 
		\includegraphics{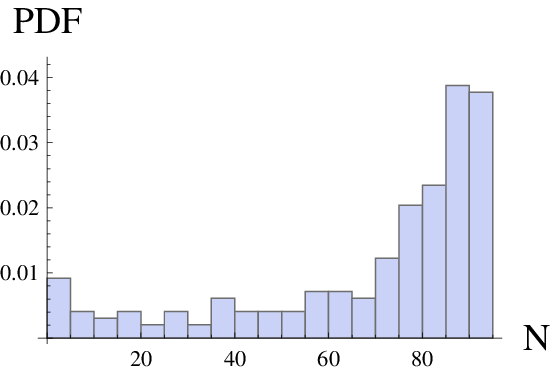}\\
		\includegraphics{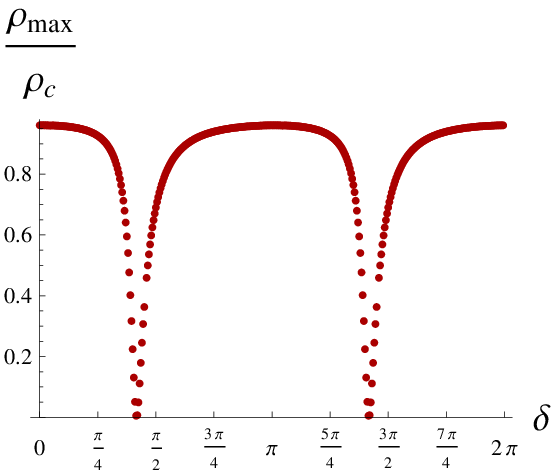}\hfill 
		\includegraphics{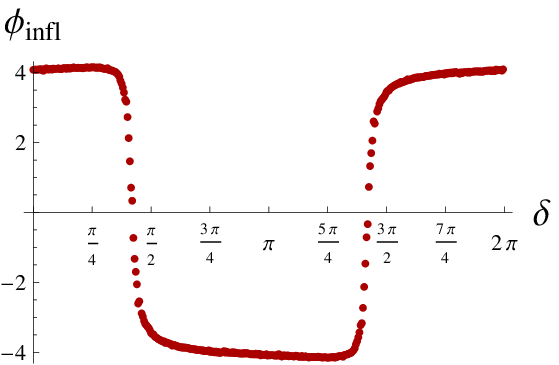}\hfill 
		\includegraphics{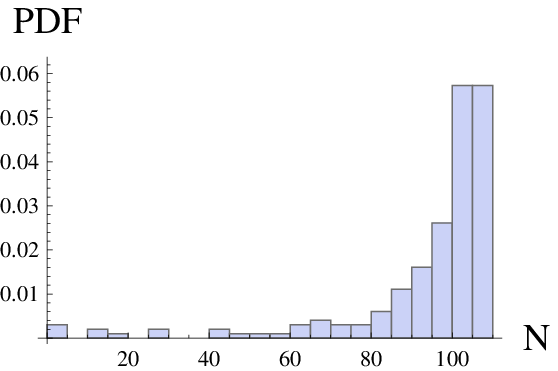}\\
		\includegraphics{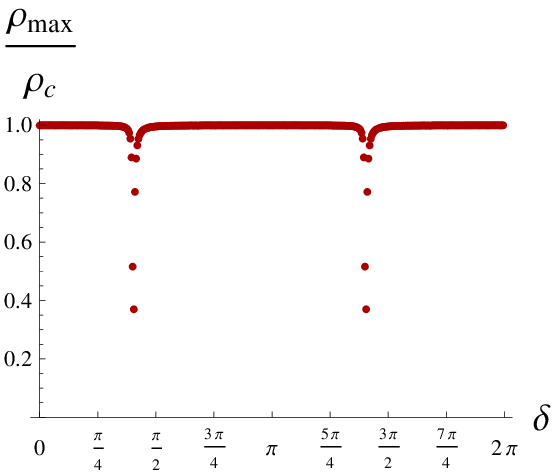}\hfill 
		\includegraphics{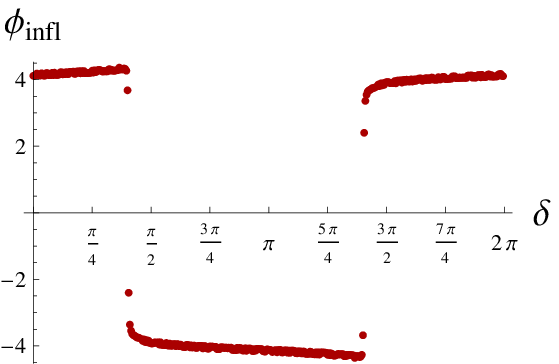}\hfill 
		\includegraphics{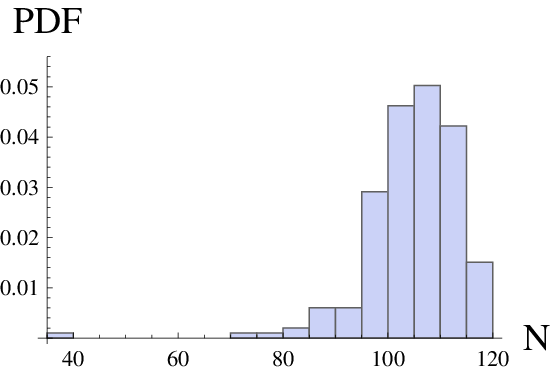}\\
		\includegraphics{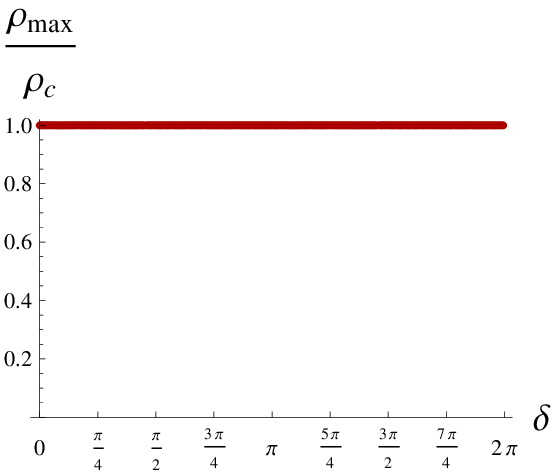}\hfill 
		\includegraphics{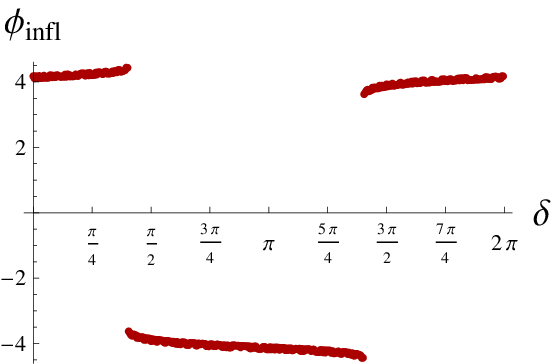}\hfill 
		\includegraphics{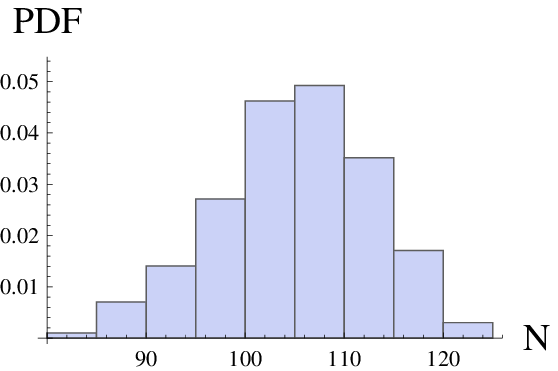}\\	
		\caption{Results of the simulations, for each row, starting from the top $\sQ(0)=\{10^{-2}\frac{\kappa}{3}\rho_0,10^{-3}\frac{\kappa}{3}\rho_0,10^{-4}\frac{\kappa}{3}\rho_0,10^{-6}\frac{\kappa}{3}\rho_0,0\}$. First column: the maximum value of $\rho$ (which is the value of $\rho$ at the bounce) normalized to $\rho_c$.  Second column: $\phi$ at the start of slow-roll inflation. Third column:  Numerically calculated probability distribution function of the number of e-folds of slow-roll inflation, given that all $\delta$ have an equal probability.}
		\label{Big}
	\end{center}
\end{figure*}

\begin{figure}
	\begin{center}
		\includegraphics[width=\columnwidth]{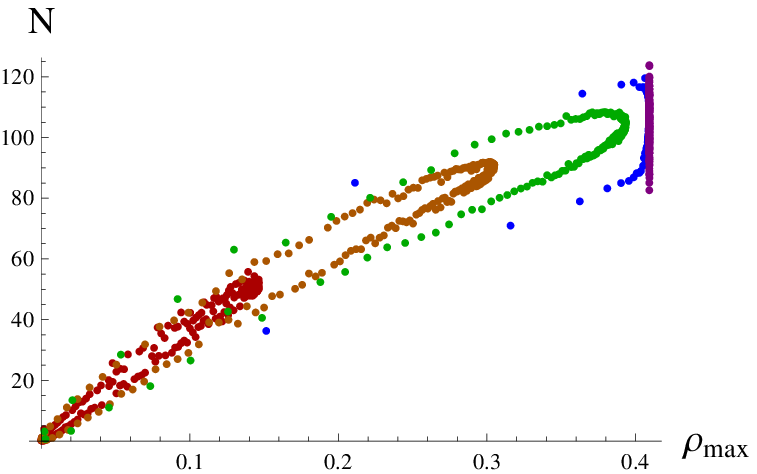}
		\caption{Number of e-folds as a function of $\rho$ at the bounce. Red: $\sQ(0)=10^{-2}\frac{\kappa}{3}\rho_0$. Orange: $\sQ(0)=10^{-3}\frac{\kappa}{3}\rho_0$. Green: $\sQ(0)=10^{-4}\frac{\kappa}{3}\rho_0$. Blue: $\sQ(0)=10^{-6}\frac{\kappa}{3}\rho_0$. Purple: $\sQ(0)=0$. Different points of the same color correspond to different values of delta}
		\label{Nrho}
	\end{center}
\end{figure}

\section{Results}
It is important to stress that our results are model dependent. The results presented in this section are only true for the specific matter content described in Section \ref{M} with $m=1.21\times 10^{-6}$. Any more general statement would require further investigations.
\\

Fig. \ref{typ} shows a typical evolution around the bounce (the bounce being defined as the point in time when $H=0$).
At the bounce, the Universe is completely dominated by the kinetic energy and the shear. The kinetic energy is very large for a very short time. This gives the scalar field a boost, and lifts it up to create the initial conditions for slow-roll inflation. 

The evolution of the scalar field $\phi$, shown in the upper panel of Fig. \ref{typ}, is not at all time symmetric around the bounce. After the bounce one can see the beginning of slow-roll inflation, with a high and almost constant value of $\phi$. But there is, in general, no analog slow-roll deflation before the bounce. This is very typical of solutions for which the initial conditions are specified in the contraction phase before the bounce. The reason for this is that in forward time evolution, slow-roll deflation is a repellant, while slow-roll inflation is an attractor. This behavior have been thoroughly studied for the isotropic case in \cite{Linsefors:2013cd}. 

As confirmed by the results given later in this section, if there is a lot of shear, the bounce happens at a lower value of the kinetic energy, and the scalar field is not lifted as high as in the isotropic case, which leads to less slow-roll inflation.
\\

The number of e-folds of slow-roll inflation, $N$, can be calculated from the value of the scalar field at the beginning of the slow-roll phase $\phi_{infl}$:
\be
N=2\pi\phi_{infl}^2,
\ee
where the field is expressed in Planck units.
The simulations confirm that  slow-roll inflation starts when $|\phi|$ reaches its maximum value after the bounce.

\subsection{Symmetry}

Classically ({\it i.e.} with no quantum corrections), the value of $S$ does not influence the evolution of the total Hubble parameter $H$ and  therefore does not have any impact on the inflation in this scenario. In effective LQC, however, the situation is different. A large asymmetry (\emph{i.e.} $|S|$ close to one) will stretch out the bounce. This can be seen in Fig. 6 of \cite{Linsefors:2013bua} and in Fig. 11 of \cite{Gupt:2013swa}.

It should be noticed that in \cite{Gupt:2013swa} the variables used for the shear are not the same as the ones chosen here. However, in \cite{Gupt:2013swa}, the initial conditions are given at the bounce, \emph{i.e.} $H(0)=0$, so that $\sQ(0)$ becomes completely fixed by \eq{fried} together with the initial values for the matter fields $\phi(0)$, $\dot\phi(0)$. Therefore, varying $\sigma^2$ in the setting of \cite{Gupt:2013swa} is equivalent to varying $S$ in the setting of this paper. 

The more dominant the shear is at the bounce, the stronger becomes this stretching effect, and vice versa. A longer bounce epoch might influence the outcome of $\phi_{infl}$ and thereby the duration of slow-roll inflation. We will therefore investigate how strong this  effect is for the range of initial conditions studied in this article.

Fig. \ref{S} shows the evolution of $H$ and $H_i$ just around the bounce for $\sQ(0)=10^{-2}\frac{\kappa}{3}\rho_0$, $\delta=0$ and some  values of $S$. One can see that in this case the evolution of $H$ is very similar for all choices of $S$.

Fig. \ref{SS} shows the resulting $\phi_{infl}$, as an output of simulations for different values of $\sQ(0)$, $\delta$ and $S$. One sees that, up to numerical errors, there is no visible dependence on $S$. We conclude that the symmetry variable has negligible effect on the duration of slow-roll inflation in the parameter range of this article. Therefore, $S=0$ is chosen for the rest of this study.

\subsection{Shear}

In this section we vary $\sQ(0)$ and $\delta$ to see how this affects the length of slow-roll inflation.

Fig. \ref{Big} shows the results of extensive simulations carried out by varying both parameters. It can be concluded that, in general, the number of e-folds goes down when the shear increases. But a greater shear will also lead to a bigger spread in the number of e-folds, depending on the initial angle $\delta$.

We stress that our results are complementary to those of \cite{Gupt:2013swa} and cannot be directly compared because the initial conditions are set in two very different manners. In this article we chose to assume that the natural place to put initial conditions in the classical contracting past, as far away as possible from the highly quantum bounce region, and in agreement with usual causality.

For a model to be compatible with the observations, about 60 e-folds or more of slow-roll inflation are needed \cite{linde}. In Fig. \ref{Big}, one can see that for the given parameters, this bound goes somewhere between $\sQ=10^{-2}\frac{\kappa}{3}\rho_0$ and $\sQ=10^{-3}\frac{\kappa}{3}\rho_0$. With the help of \eq{prop2}, one can express this in a slightly more general way. As long as \eq{cond} holds,
\be
\sQ(t) = \frac{\sQ(0)}{\rho_0^2} \left<\rho(t)\right>^2.
\ee
To have a long enough slow-roll inflation period we need $\sQ(0) \lesssim 5*10^{-3}\frac{\kappa}{3}\rho_0$ for $\frac{\kappa}{3}\rho_0=10^{-3}m^{2}$, which is equivalent to
\ba
\sQ(t) &\lesssim& \frac{(5*10^{-3})(10^{-3}m^{2})}{(\frac{3}{\kappa}10^{-3}m^{2})^2} \left<\rho(t)\right>^2\\\
&=& \frac{5}{m^2} \left(\frac{\kappa}{3}\left<\rho(t)\right>\right)^2.\nn
\ea
during the early evolution.
\\

Fig. \ref{Nrho}, shows the same data as Fig. \ref{Big} but in a different way. Here one can clearly see that the number of e-folds of slow-roll inflation depends strongly on  $\rho_{max}$, which is of course not surprising. When the shear vanishes, $\rho_{max}$ becomes fixed, otherwise it can vary between zero and the associated maximum value.

From Fig. \ref{Nrho} one can reed out that to get more than 60 e-folds of inflation, $\rho_{max}\gtrsim0.18$ is required. However, this result is highly dependent on the fact that the initial conditions are set in the past, long enough before the bounce. As pointed out earlier, using initial conditions in the contracting phase promotes a certain kind of solutions. This is precisely the spirit of this study and of the previous ones \cite{Linsefors:2013bua,Linsefors:2013cd} that we intend to generalize here.

\begin{figure}
	\centering
		\includegraphics[width=\columnwidth]{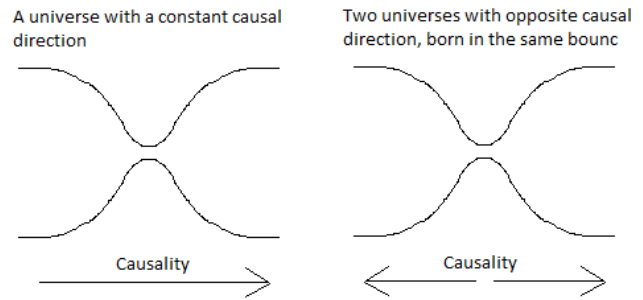}
	\caption{Two possibilities for the causal structure of a bouncing universe.}
	\label{kaos}
\end{figure}

\section{Discussion}\lb{dis}

In this work we have investigated how the amount of shear affects the duration of slow-roll inflation in effective LQC, when the initial conditions are set in the far past. We have found that a larger shear will in general lead to a lower energy density at the bounce. 

At the bounce, the energy density is dominated by the kinetic energy. For a very short time around the bounce, of the order of a Planck time, the kinetic energy is of the order of the Planck energy. This peak of kinetic energy will lift the potential energy up to a value of the order $m^2$. When the peak of kinetic energy has died down, only the potential energy remains, and the slow-roll inflation starts. 

However, the bounce will occur when $\rho$ together with $\sQ$ are big enough so that the left hand side of \eq{fried} vanishes. Naturally, more shear will leave less room for the energy density. The growth of kinetic energy is interrupted at a lower value than in the isotropic case, leading to a smaller boost of the potential energy, leading to less slow-roll inflation. It should also be noticed that the initial parameter $\delta$, relating the amount of initial kinetic and potential energy, has an impact on the duration of slow-roll inflation. However, $\delta$ should be seen as a truly random parameter, and we should therefore consider the results for a flat statistical distribution of $\delta$.

We conclude that to have a good probability for at least 60 e-folds of inflation in this model, the initial conditions must fulfill: $\sQ\lesssim\frac{5}{m^2}\left(\frac{\kappa}{3}\rho_0\right)^2.\nn$\\

There are two major schools on how to set initial conditions in LQC. One is, as in this study, to set the initial conditions in the remote past. The other direction is to set the initial conditions at the bounce. Which way is correct is a rather philosophical question. Causality in our universe seems to be running in a certain direction, which we call forward in time. When simulating almost any epoch in the Universe, physicists always run the simulations in the forward time direction, even though the laws of physics are time symmetric. The reason for this is because we are not actually interested by {\it all} possibilities, we are only interested in the most probable ones. The direction of causality is important for probabilities, {\it e.g.} an attractor in one time direction will be a repellant in the opposite time direction. The question of when to set the initial conditions in the simulation is an important one.

What is then the correct initial time in LQC? Well, it depends on how one thinks about the bounce. If one believes in a constant causal direction, as in the left panel of Fig. \ref{kaos}, it is natural to set the initial conditions well before the bounce. However, we do not know how time behaves in quantum gravity, so this might not be the case. An alternative history is sketched in the right panel of Fig. \ref{kaos}, where two universes with opposite causal direction are born from the same "bounce". In this case, it would be natural to assign initial conditions at the bounce. We do not know which, if any, of these two cases is correct. It is therefore reasonable to consider both paths. Anyway, each of them has its own problem with anisotropies.

In this article, we have investigated the case of initial conditions set well before the bounce. We have found that if the shear is too large when compared to the energy density, we will not have enough inflation to explain the known universe. This means that the shear "decreases" the amount of solutions that are compatible with data, and therefore the "naturalness" of the model. But on the other hand, this means that if the model is correct, the number of e-folds is probably not much larger that the lower bound required by observations. This is good for phenomenology as the quantum gravity effects may then not have been fully washed out by inflation.

The case of initial conditions  set at the bounce was investigated in \cite{Gupt:2013swa} where it was found that the anisotropies are not a problem for inflation. However, only solutions leading to a classical limit in the future where considered. But the vast majority of solutions do never lead to anything remotely like a classical universe at all \cite{Linsefors:2013bua}. When starting from a classical contracting universe, the nice solutions ({\it i.e.} the ones leading to a classical expanding universe after the bounce) are automatically singled out. But if initial conditions are put at the bounce, something else is needed to play this role.


\begin{thebibliography}{99}

\bibitem{rovelli1} 
P.~Dona \& S.~Speziale, arXiv:1007.0402V1;\\
A.~Perez, arXiv:gr-qc/0409061v3;\\
R. Gambini \& J. Pullin, {\it A First Course in Loop Quantum Gravity}, Oxford, Oxford University Press, 2011;\\
C.~Rovelli, arXiv:1102.3660v5 [gr-qc];\\
C.~Rovelli, {\it  Quantum Gravity}, Cambridge, Cambridge University Press, 2004;\\
C.~Rovelli, Living Rev. Relativity, 1, 1, 1998;\\
L.~Smolin, arXiv:hep-th/0408048v3;\\
T.~Thiemann, Lect. Notes Phys., 631, 41, 2003; T.~Thiemann, {\it Modern Canonical Quantum
General Relativity}, Cambridge, Cambridge University Press, 2007

\bibitem{lqc_review} 
A.~Ashtekar, M.~Bojowald, and J.~Lewandowski, Adv.\ Theor.\ Math.\ Phys.\  7, 233, 2003;\\
A.~Ashtekar, Gen. Rel. Grav. 41, 707, 2009;\\
A.~Ashtekar, P. Singh, Class. Quantum Grav. 28, 213001, 2011;\\
M.~Bojowald, Living Rev. Rel. 11, 4, 2008;\\
M.~Bojowald, arXiv:1209.3403 [gr-qc];\\
K. Banerjee, G. Calcagni, and M. Martin-Benito, SIGMA 8, 016, 2012;\\
G. Calcagni, Ann. Phys. (Berlin) 525, 323, 2013;\\
I. Agullo and A. Corichi, in "The Springer Handbook of Spacetime," edited by A. Ashtekar and V. Petkov. (Springer-Verlag, at Press); arXiv:1302.3833 [gr-qc];\\
A. Barrau {\it  et al.}, Class. Quant. Grav. 31, 053001, 2014

\bibitem{diener} 
P. Diener, B. Gupt, P. Singh, arXiv:1402.6613

\bibitem{mubar} 
A. Ashtekar, E. Wilson-Ewing, Phys. Rev. D 79, 083535, 2009

\bibitem{bianchiI_lqc} 
D. Cartin, G. Khanna, Phys. Rev. Lett. 94, 111302, 2005;\\
G. Date, Phys. Rev. D 72, 067301, 2005;\\
D. Cartin, G. Khanna, Phys. Rev. D 72, 084008, 2005;\\
D.-W. Chiou, Phys. Rev. D 75, 024029, 2007;\\
D.-W. Chiou, arXiv:gr-qc/0703010;\\
D.-W. Chiou, Kevin Vandersloot, Phys. Rev. D, 76, 084015, 2007;\\
D.-W. Chiou, Phys. Rev. D 76, 124037, 2007;\\
L. Szulc, Phys. Rev. D 78, 064035, 2008;\\
M. Martin-Benito, G.A. Mena Marugan, T. Pawlowski, Phys. Rev. D 78, 064008, 2008;\\
D.-W. Chiou, arXiv:0812.0921 [gr-qc];\\
P. Dzierzak, W. Piechocki, Phys. Rev. D 80, 124033, 2009; \\
M. Martin-Benito, G. A.Mena Marugan, Tomasz Pawlowski, Phys. Rev. D80, 084038, 2009;\\
P. Dzierzak, W. Piechocki, Annalen Phys. 19, 290, 2012; \\
P. Malkiewicz, W. Piechocki, P. Dzierzak, Class. Quant. Grav. 28, 085020, 2011;\\
M. Martin-Benito, L.J. Garay, G.A. Mena Marugan, E. Wilson-Ewing, J. Phys. Conf. Ser. 360, 012031, 2012;\\
V. Rikhvitsky, B. Saha, M. Visinescu, Astrophys. Space Sci. 339, 371, 2012;\\
P. Singh, Phys. Rev. D 85, 104011, 2012;\\
F. Cianfrani, A. Marchini, G. Montani, Europhys. Lett. 99, 10003, 2012;\\
K.Fujio, T. Futamase, Phys. Rev. D 85, 124002, 2012; \\
P. Singh, J. Phys. Conf. Ser. 360, 012008, 2012;\\
X. Liu, F. Huang, J.-Y. Zhu, Class. Quant. Grav. 30, 065010, 2013;\\
X.-J. Yue, J.-Y. Zhu, arXiv:1302.1014 [gr-qc]



\bibitem{Linsefors:2013bua}
  L.~Linsefors and A.~Barrau,
  Class. Quant. Grav. 31, 015018, 2014

\bibitem{Linsefors:2013cd}
  L.~Linsefors and A.~Barrau,
  Phys. Rev. D 87, 123509, 2013
	
\bibitem{Gupt:2013swa}
  B.~Gupt and P.~Singh,
  Class. Quant. Grav.  30, 145013, 2013 
  




\bibitem{linde}
A. Linde, 	arXiv:1402.0526

\end{thebibliography}
\end{document}